\documentclass[letterpaper,english,showpacs,preprintnumbers,amsmath,amssymb,floatfix,superscriptaddress,showkeys,prl,preprint]{revtex4}
\usepackage[T1]{fontenc}
\usepackage[latin1]{inputenc}
\usepackage{amsmath}
\usepackage{graphicx}
\usepackage{amssymb}

\makeatletter
\usepackage{graphicx}

\usepackage{dcolumn}
\usepackage{bm}
\usepackage{times}

\usepackage{babel}

\usepackage{babel}
\makeatother
\begin{document}

\title{Helical structures from an isotropic homopolymer model}

\author{James E. Magee}

\email{j.magee@manchester.ac.uk}

\affiliation{School of Chemical Engineering and Analytical Science, The University
of Manchester, PO Box 88, Sackville Street, Manchester M60 1QD, United
Kingdom}

\author{Victor R. Vasquez}

\affiliation{Chemical Engineering Department, University of Nevada at Reno, Reno,
NV 89557, USA }

\author{Leo Lue}

\email{leo.lue@manchester.ac.uk}

\affiliation{School of Chemical Engineering and Analytical Science, The University
of Manchester, PO Box 88, Sackville Street, Manchester M60 1QD, United
Kingdom}

\date{\today{}}

\begin{abstract}
We present Monte Carlo simulation results for square-well homopolymers
at a series of bond lengths. Although the model contains only isotropic
pairwise interactions, under appropriate conditions this system shows
spontaneous chiral symmetry breaking, where the chain exists in either
a left- or a right-handed helical structure. We investigate how this
behavior depends upon the ratio between bond length and monomer radius. 
\end{abstract}



\maketitle
\label{sec:intro}Naturally occurring polymers, such as proteins (polypeptides)
and nucleic acids (polynucleotides), are capable of adopting very
specific single-chain conformations under certain conditions. It is
believed that this structural specificity plays a crucial role in
the biological function of a given molecule. Consequently, the essential
mechanism of how the overall structure of a polymer arises from the
basic interactions between its constituent monomers is of great fundamental
interest. A recurring motif in many diverse biopolymers is the helix.
This structure has also been observed in synthetic polymers --- for
example, meta-phenylacetylene oligomers \cite{phenylacetylene}, and
the isotactic form of poly(methyl methacrylate) \cite{Kusanagi76}
in various solvents. The occurrence of helical structures in polymers
with substantially different monomer chemistries suggests an underlying
generic behavior, which has prompted significant efforts to produce
{}``minimal'' models that naturally display helical structures \cite{Muthukumar1996,KempChen1998,BuhotHalperin2002,Varshneyetal2004}.
An underlying assumption in the vast majority of these models is stated
explicitly by Kemp and Chen \cite{KempChen2001}: {}``an isotropic
potential interaction is not sufficient to produce helical ground
states.'' In this letter, we present simulation results for a polymer
model using only pairwise isotropic monomer interactions. We explicitly
demonstrate that this model breaks chiral symmetry at low temperatures
and show evidence that it forms stable helices.

The general theoretical perspective \cite{ZimmBragg1959,LifsonRoig1961}
has been to view helix formation as a discrete process. Each polymer
element (monomer) can only take on one of two states, {}``helical''
or {}``coil'' (non-helical), and a free energy is assigned to each
of these states. This approach has been fairly successful in describing
the helix-coil transition in polymer systems. One shortcoming of this
approach, however, is that it presupposes the existence of a particular
helix, and, as a consequence, it is difficult to determine the conditions
under which a helical conformation can naturally arise.

Another line of research has been to apply molecular simulations techniques
to examine the structural behavior of minimal polymer models, which
has lead to interesting results. Zhou and coworkers discovered \cite{Zhouetal1996,Zhouetal1997}
a freezing transition for linear homopolymers composed of tangent-hard
spheres that interact with each other through an attractive square-well
potential. During this discontinuous transition, the collapsed {}``molten''
globule state freezes into a highly symmetric crystalline-like state.
While this transition does result in structural specificity, the observed
ordered state does not resemble any of the structural motifs seen
in biopolymers. Noguchi and Yoshikawa \cite{NoguchiYoshikawa1998}
extended the model of Zhou and coworkers by adding a bond bending
potential, which introduced stiffness to the polymer model. Upon increasing
the stiffness of the chain, toroidal and rod-like phases were observed
--- analogous to structures exhibited by isolated DNA chains \cite{Bloomfield1996,Plumetal1990}.
However, they did not observe the formation of helices in their simulation
studies. Kemp and Chen \cite{KempChen1998,KempChen2001} have developed
a simple homopolymer model that displays a helix-coil transition;
however, the monomer-monomer interactions in this model are asymmetric.

For thick strings confined within a fixed geometry (e.g., inside a
cube), a helix can be the optimal packing configuration \cite{Maritanetal2000}.
These {}``close-packed'' configurations may play a role in determining
the structure of linear chains in a similar manner as the close-packed
structures for hard spheres play for the solid phase of spherically
symmetric molecules. Structures other than the helix were shown to
optimize the packing under appropriate conditions; the nature of the
optimal structure depends on such parameters as the thickness and
length of the string, as well as the geometry of the confinement.

Although anisotropic, attractive interactions between monomers can
induce a chain to form a helix, we will demonstrate that the delicate
interplay between purely isotropic, attractive interactions and packing
(due to excluded volume interactions) of monomer segments can also
induce a helix. A homopolymer chain in poor solvent conditions will
tend to collapse into a compact conformation in order to maximize
self-interactions. Beyond a critical packing fraction, optimal configurations
tend to maximize the entropy available to a system. Therefore, for
situations where the helix is the optimal configuration for a string,
we expect a transition to a helical structure for a sufficiently string-like
polymer model.

\label{sec:simul}We consider polymers consisting of $N$ linearly
bonded hard spheres of diameter $\sigma$. The distance between bonded
spheres are restricted to lie between $0.9l$ and $1.1l$, where $l$
is the bond length, which is not necessarily equal to $\sigma$. Spheres
that are not directly bonded interact via a square-well potential
$u(r)$, given by: \begin{eqnarray}
u(r) & = & \left\{ \begin{array}{rl}
\infty & \mbox{for $r<\sigma$}\\
-\epsilon & \mbox{for $\sigma<r<\lambda\sigma$}\\
0 & \mbox{for
$\lambda\sigma<r$}\\
\end{array}\right.\label{eq:SWpotn}\end{eqnarray}
 where $r$ is the distance between the centers of the spheres, $\lambda\sigma$
is the width of the attractive square-well ($\lambda=1.5$ in this
work), and $\epsilon$ is the magnitude of the interaction, which
sets the energy scale of the model. For $l=\sigma$, this model is
precisely the same as that studied by Zhou and coworkers \cite{Zhouetal1996,Zhouetal1997}.

We have carried out Monte Carlo simulations to study the behavior
of unconfined, isolated polymers with $N=20$ (20-mers) and values
of $\sigma/l$ between $1.0$ and $1.9$. Parallel tempering \cite{SwendsenWang1996}
was employed, with 16 replica systems at temperatures between $k_{B}T/\epsilon=0.2$
and $k_{B}T/\epsilon=7.0$ (where $k_{B}$ is the Boltzmann constant),
placed such that the ratio between successive temperatures was constant
\cite{Kofke2002}. Within each replica system, the configuration of
the polymer was modified using a combination of pivot moves \cite{MadrasSokal1988},
kink-jump moves \cite{Sadus2002}, and continuum configurational bias
regrowth moves \cite{EscobedodePablo1995}. Bond length fluctuations
were implemented by performing standard particle displacement moves
\cite{AllenandTildesley1987} on individual monomer spheres. After
$400$ attempted moves in each box, replica exchange moves were attempted.
Samples were taken every $80$ attempted swap moves. The simulation
consisted of a total of $8\times10^{5}$ attempted swap moves. Smooth
interpolation of data between the replica temperatures has been performed
using multiple histogram reweighting \cite{FerrenbergSwendsen1989}.

The structure of the polymer conformations has been characterized
using the local torsion $\tau_{i}$ and curvature $\kappa_{i}$ \cite{Kreyszig}
around the bond vector joining monomers index $i$ and $i+1$ ($i\in\left[1,N\right]$).
Curvature is defined so that is always positive; the sign of the torsion
around a bond indicates whether it has a left- or right-handed {}``twist''.
These quantities are calculated using a central difference scheme
\cite{Chapra98}; as such, $\tau_{i}$ is defined on $i\in\left[3,N-2\right]$,
and $\kappa_{i}$ on $i\in\left[2,N-1\right]$. It is a necessary
and sufficient condition for a perfect helical structure to have $\tau_{i}/\kappa_{i}$
independent of $i$ \cite{Struik}, taking the value $p/2\pi r_{h}$,
where $p$ is the pitch of the helix and $r_{h}$ is the radius. This
quantity will be quoted as an average along the polymer for a given
conformation, denoted $\gamma$, and given by: \begin{equation}
\gamma={\displaystyle \sum_{i=3}^{N-2}}\left(\tau_{i}/\kappa_{i}\right)/\left(N-4\right)\label{eq:configavg}\end{equation}

\noindent This conformation average should not be confused with the
usual thermal (ensemble) average. As an indicator of {}``helicity'',
we also measure the conformational root variance for this average,
$s_{\gamma}$. This quantity should take consistently low values for
a helical phase, since the ratio $\tau_{i}/\kappa_{i}$ will be constant
for a perfect helical configuration. 

\begin{figure}
\begin{center}\includegraphics{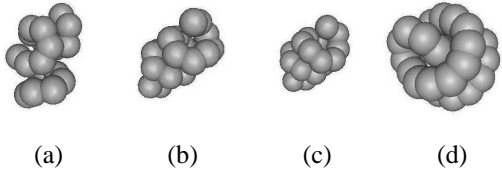}\end{center}

\caption{\label{configs:fig} Snapshot homopolymer configurations (a) 20-mer
with $\sigma/l=1.6$ and $k_{B}T/\epsilon=0.75$ (globule phase),
(b) 20-mer with $\sigma/l=1.6$ and $k_{B}T/\epsilon=0.48$ (helix
1 phase), (c) 20-mer with $\sigma/l=1.6$ and $k_{B}T/\epsilon=0.2$,
(helix 2 phase) and (d) 32-mer with $\sigma/l=1.9$ and $k_{B}T/\epsilon=0.2$.}
\end{figure}

\label{sec:res}%
\begin{figure}
\begin{center}\includegraphics{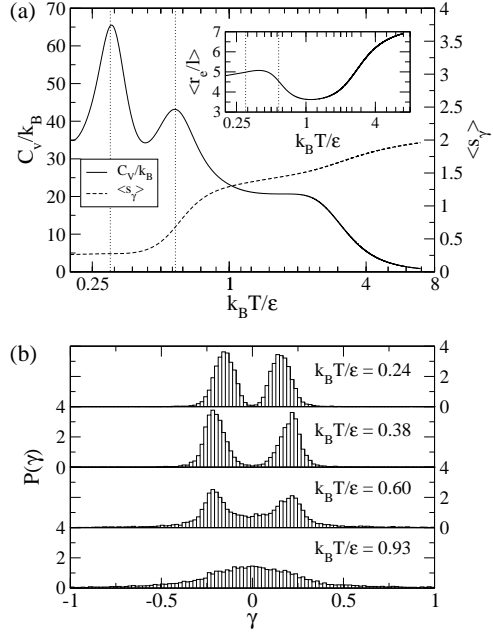}\end{center}

\caption{\label{cv:fig} (a) Variation with temperature of the isochoric heat
capacity $C_{v}/k_{B}$ and mean conformational root variance, $\left\langle s_{\gamma}\right\rangle $
for a homopolymer composed of 20 monomers, $\sigma/l=1.6$; jackknife
estimated error is never more than 2\% of the values shown. (inset)
Mean end-to-end distance $\left\langle r_{e}\right\rangle $ for the
same system; jackknife estimated error is never more than 2\% of the
values shown. Dotted drop lines indicate the temperatures of maxima
in $C_{v}$. (b) Probability density histograms $P(\gamma)$ for the
ratio of configuration average torsion and curvature for the same
system, temperatures as in legend. }
\end{figure}
The variation of the isochoric heat capacity $C_{v}$ with temperature
$T$ for a typical 20-mer chain ($\sigma/l=1.6$) is shown in Fig.~\ref{cv:fig}(a).
Cooling from high temperature, the heat capacity rises through a broad
peak or shoulder as the radius of gyration (not shown) undergoes a
large, though gradual, decrease. This is the {}``$\Theta$-point''
transition from extended coil to disordered globule, where the repulsive
excluded volume interactions between the monomers are balanced by
the attractive interactions. This transition is analogous to the vapor-liquid
transition in bulk fluids. A snapshot configuration from the globule
phase is shown in Fig.~\ref{configs:fig}(a).

Upon further cooling, the heat capacity passes through another peak
at $k_{B}T/\epsilon=0.58$. We attribute this peak to the adoption
of a compact, ordered structure by the polymer and denote this temperature
$T_{f}$. Snapshot configurations taken from the simulations below
this temperature appear to be helical in character. An example is
given in Fig.~\ref{configs:fig} (b). Since snapshot configurations
are not a reliable indicator of structure, we have characterized the
observed structures using the conformational mean torsion-curvature
ratio $\gamma$, and its conformational root variance $s_{\gamma}$.

Probability density plots $P(\gamma)$ for this system are shown in
Fig.~\ref{cv:fig} (b). At high temperature, these have an approximately
Gaussian form, centered on zero. As the system is cooled below $T_{\theta}$,
$P(\gamma)$ gradually becomes double peaked, with the two peaks symmetrically
placed around zero and separated by a shallow trough. The free energy
barrier between states with $\gamma>0$ (right-handed) and $\gamma<0$
(left-handed) is relatively low, and the polymers can freely exchange
between them. There is no clear thermodynamic indicator of the temperature
below which this characteristic appears. 

Below $T_{f}$, the character of $P(\gamma)$ changes once again.
As the temperature passes below $T_{f}$, the two symmetrically placed
peaks become separated by a region of very low probability. This marks
the breaking of chiral symmetry, with the system adopting either a
counterclockwise ($\gamma<0)$ or a clockwise ($\gamma>0$) helix.
These two states are separated by a large free energy barrier, and,
consequently, the polymers are not expected to exchange between the
counterclockwise and clockwise helix conformations. It should be noted
that sampling of both peaks occurs mainly because of the use of parallel
tempering in the simulations. That the two peaks are sampled with
approximately equal weight is evidence that the simulations are ergodic
and well equilibrated. Passing down in temperature through $T_{f}$,
a sigmoidal decrease in $\left\langle s_{\gamma}\right\rangle $ to
a low, approximately temperature-independent value is also observed,
showing that the $\tau/\kappa$ values for individual bonds are becoming
strongly correlated, indicating a transition to a helical state. %

It can be seen from Fig.~\ref{cv:fig} (a) that there is a further
low-temperature peak in $C_{v}$ at $k_{B}T/\epsilon=0.3$. Upon cooling
across this temperature, $P\left(\gamma\right)$ remains double peaked
and shows a sudden shift inward, toward zero. Snapshot configurations
remain helical, with an example shown in Fig.~\ref{configs:fig}
(c). The inset to Fig.~\ref{cv:fig} (a) shows that the mean end-to-end
difference for the polymer decreases upon cooling across this temperature.
This tallies with the observed decrease in $\gamma,$ indicating a
decrease in pitch and/or increase in radius for a helical configuration.

\begin{figure}
\begin{center}\includegraphics{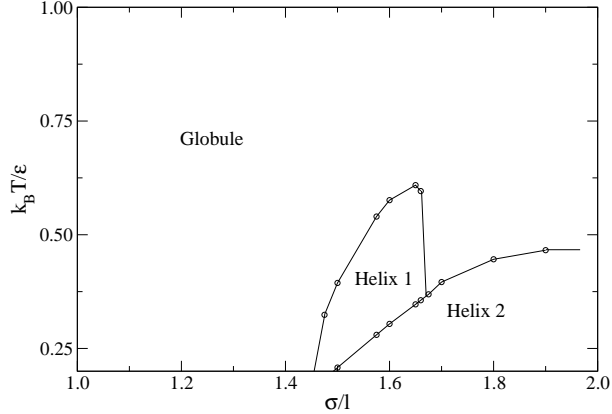}\end{center}

\caption{\label{phasediag:fig} Schematic phase diagram. Points represent
observed isochoric heat capacity peaks; the statistical uncertainty
in the positions is smaller than symbol size. Lines serve only as
guides to the eye; see text for discussion. Text indicates the structure
observed in that region. }
\end{figure}

The general, qualitative behavior shown in Fig.~\ref{cv:fig} is
also observed for polymers with $\sigma/l=1.475$ to $\sigma/l=1.65$.
For values $\sigma/l\geq1.7$, the behavior is as described above,
but with only a single low-temperature peak observed in the isochoric
heat capacity. For values $\sigma/l<1.475$, chiral symmetry breaking
is not observed; the observed histograms for $P(\gamma)$ remain unimodal
and centered on zero across the studied temperature range. These observations
are summarized in the schematic phase diagram presented Fig.~\ref{phasediag:fig}.%

\label{sec:Discussion}We have shown that stiffness of the polymer
induced by the bulkiness of the monomers plays an important role in
the development of structure at low temperatures. For values of $\sigma/l\geq1.475$,
the equilibrium low temperature structure appears to be helical (see
Fig.~\ref{phasediag:fig}); further evidence for this is provided
by the consistently low values for the quantity $\left\langle s_{\gamma}\right\rangle $
under these conditions, indicating an approximately constant value
for $\tau_{i}/\kappa_{i}$ across individual conformations. The torsional
symmetry breaking associated with helix structure formation must be
associated with a thermodynamic phase transition for an infinite system;
no attempt is made to establish the order of this transition, as the
small size of the system studied would make any such attribution unreliable.
We have identified these transitions by the associated increase in
energy fluctuations (maxima in heat capacity). Across the temperature
range studied, no ordered low temperature structures are observed
for $\sigma/l<1.475$. For $\sigma/l=1$, the globule should freeze
into an ordered crystalline phase \cite{Zhouetal1996,Zhouetal1997}
at sufficiently low temperature. For other values of $\sigma/l$,
however, the low temperature structure is still unclear.

For values $1.475\leq\sigma/l\leq1.66$, the helix-globule transition
line is reentrant. With increasing $\sigma/l$, the helix initially
gains, then rapidly loses stability with respect to the globule. At
low temperature, there is a line of heat capacity maxima which coincides
with a sudden drop in the torsion-curvature ratio and decrease in
mean end-to-end distance for the observed helices. We interpret it
as indicating a transition to a less extended helical structure at
low temperature, since $\tau/\kappa$ gives the pitch-radius ratio
for a helix. This line of maxima appears to join with the helix-globule
transition line for $\sigma/l\geq1.675$. No further heat capacity
maxima or discontinuities in torsion are observed between this line
and the lowest temperature studied. Since the high temperature helix
is not observed for $\sigma/l\geq1.675$, and there must exist a transition
line between a symmetry breaking and a non-symmetry breaking state,
we can therefore infer the existence of a helix-helix-globule triple
point in the range $1.66\leq\sigma/l\leq1.675$. The less extended
helical structure is the low-temperature equilibrium configuration
observed for all $\sigma/l\geq1.475$, and appears to increase in
stability monotonically with $\sigma/l$. Extrapolation of the helix-globule
transition line and the helix-helix transition line to low temperatures
for $\sigma/l<1.475$ suggests that a further, low temperature helix-helix-globule
triple point may exist. 

The precise reasons for the occurrence of two helical structures separated
by an apparent transition line, and for the sudden decrease in stability
of the high temperature helix are, as yet, unclear. Results using
an anisotropic model \cite{KempChen1998,KempChen2001}, constructed
to yield a specific helix, suggest a two-step mechanism for helix
formation, where the initial globule-helix transition results in a
structure with disordered end segments, followed by a transition to
a fully-helical structure. Another speculative possibility is that
allowed helical states are discretized due to a preference for monomers
to sit at interstitial sites in neighbouring layers of the helix,
leading to a succession of helical phases separated by transitions
rather than a continuum of possible torsion values. Further study
will be necessary to understand these behaviors. 

Another factor playing an important role is the overall \emph{length}
of the polymer. We have studied the square-well chain model for different
lengths $N$, and preliminary results suggest that, though the system
breaks chiral symmetry at low temperature for all chain lengths studied,
helices are only stable inside a window of values for $N$. An example
low-temperature snapshot configuration for a longer polymer is shown
in Fig.~\ref{configs:fig} (d); a detailed examination of the influence
of chain length on structure will be the subject of a further communication.

In summary, we have demonstrated spontaneous chiral symmetry breaking
using a model which is achiral (no dependence upon bond torsions)
and isotropic (no dependence upon bond angles). This \emph{spontaneous}
symmetry breaking reflects the formation of helical structures at
low temperature, providing support for the notion of helices as {}``optimal
configurations'' for string-like objects \cite{Maritanetal2000},
which as such do not need to be {}``engineered in'' to a model or
molecule through specific directional interactions. We do not seek
to suggest that directional interactions do not play an important
role in the formation and stabilisation of helical structure; however,
the purely steric effect which we have demonstrated may also play
a crucial role which we have not yet been aware of. The observed phase
behavior for the model is rich, including an apparent helix-helix
transition, with a reentrant helix-globule coexistence curve for one
helix leading to a helix-globule-helix triple point, and offers many
possibilities for further investigation.

\section*{\label{sec:acknow}Acknowledgments}

Partial support for this work was provide by the BBSRC (grant reference
B17005). The authors would like to thank A.~Saiani and M.~Williams
for bringing the helical structure of PMMA to our attention.

\bibliographystyle{apsrev}

\end{document}